\documentclass[runningheads]{svmult}

\usepackage{makeidx}   % allows index generation
\usepackage{graphicx}  % standard LaTeX graphics tool
\usepackage{subeqnar}  % subnumbers individual equations
\usepackage{multicol}  % used for the two-column index
\usepackage{physprbb}  % modified textarea for proceedings,
\makeindex             % used for the subject index
\usepackage{natbib209}
\def\lesssim{\mathrel{\hbox{\rlap{\hbox{\lower4pt\hbox{$\sim$}}}\hbox{$<$}}}}
\def\gtrsim{\mathrel{\hbox{\rlap{\hbox{\lower4pt\hbox{$\sim$}}}\hbox{$>$}}}}

\newcommand{\SFRD}{\mbox{$\dot{\rho}_{\rm SFR}$}}
\newcommand{\SFRDz}{\mbox{$\dot{\rho}_{\rm SFR}(z)$}}
\newcommand{\tausf}{\mbox{$\tau_{\rm SF}$}}
\newcommand{\fesc}{\mbox{$f_{\rm esc}$}}

\def\harvarditem{\@ifnextchar[{\@harvarditem}{\@harvarditem[]}}
\def\@harvarditem[#1]#2#3#4%
    {\if\relax#1\relax\bibitem[#2(#3)]{#4}\else
        \bibitem[#1(#3)#2]{#4}\fi }

\begin{document}
\title*{Lyman Break Galaxies in the NGST Era}
\toctitle{Lyman Break Galaxies in the NGST Era}
% allows explicit linebreak for the table of content
%
%
\titlerunning{Lyman Break Galaxies in the NGST Era}
% allows abbreviation of title, if the full title is too long
% to fit in the running head
%
\author{Henry C. Ferguson\inst{1}
\and Mark Dickinson\inst{1}
\and Casey Papovich\inst{1,2}
}
\authorrunning{H.C. Ferguson et al.}
% if there are more than two authors,
% please abbreviate author list for running head
%
%
\institute{Space Telescope Science Institute, 3700 San Martin Drive,
Baltimore, MD 21218
\and 
Steward Observatory, University of Arizona, 933 North Cherry Avenue,
Tucson, AZ 85721
}

\maketitle              % typesets the title of the contribution

\begin{abstract}
With SIRTF and NGST in the offing, it is interesting to examine what the
stellar populations of $z \approx 3$ galaxies models imply for the 
existence and nature of
Lyman-break galaxies at higher redshift.  To this end, we ``turn back
the clock'' on the stellar population models that have been fit to
optical and infrared data of Lyman-break galaxies at $z \approx 3$ .
The generally young ages (typically $10^{8 \pm 0.5} \,{\rm yr})$ of these
galaxies imply that their stars were not present much
beyond $z=4.$ For smooth star-formation histories $SFR(t)$ and Salpeter
IMFs, the ionizing radiation from early star-formation in these
galaxies would be insufficient to reionize the intergalactic medium at
$z \approx 6,$ and the luminosity density at $z \approx 4$ would be
significantly lower than observed. We examine possible ways
to increase  the global star-formation rate 
at higher redshift without violating the stellar-population constraints
at $z \approx 3.$
\end{abstract}

\section{Introduction}
Near-infrared photometry provides access to the rest-frame optical
portion of  Lyman-break galaxy spectra; studies are beginning to
provide some insight into the evolutionary status of LBGs. The stellar
masses are reasonably well constrained at $M^* = 3 \times 10^{10}
M_\odot \pm 0.5 \rm dex$ \cite{PDF01,SSADGP01,Yamada02}.
These stellar masses are similar to estimates of the
masses from kinematics \cite{PSSCDMAG01,Moorwood02},
and together they suggest that Lyman break galaxies must
grow substantially if they are to become $L^*$ galaxies by redshift
$z=0$. Other stellar-population parameters such as ages, metallicities,
extinction, star- formation rates and star-formation timescales are
very poorly constrained by the fits to the spectral-energy
distributions, leaving room for a variety of evolutionary paths both
prior to and after $z \sim 3$.

At some point above redshift $z \sim 6$ the intergalactic medium was
reionized. It is not known if the sources of ionization were stars or
quasars, or whether the stars responsible for the reionization had a
mass function at all similar to that observed in the Milky Way. It is
also unclear whether reionization itself had a major role in regulating
subsequent galaxy formation, for example by suppressing star formation
in low-mass galaxy halos. As new facilities such as SIRTF and NGST come
on line, understanding the causes and effects of reionization will be
one of the major goals. With that in mind, we shall briefly review the
capabilities of  the Space Infrared Telescope Facility (SIRTF), the
Hubble Space Telescope Advanced Camera for Surveys (HST ACS), and the
Next Generation Space Telescope for studying Lyman break galaxies. We
will then look at the stellar population parameters of the $z \sim 3$
samples of Lyman-break galaxies and attempt to turn the clock back to
predict their star-formation rates at higher redshift. We find a 
somewhat surprising result: that the models imply
a luminosity density at $z = 4$ significantly below that observed,
and fail to produce enough ionizing photons at $z = 6$ to account
for reionization. We discuss modifications of the models that might
be needed to avoid these problems.

\section{Lyman-Break Galaxies in the GOODS Era}

When the HDF-N/WFPC2 observations were made, $z \approx 3$ was the
frontier for galaxy surveys.  Today it stands at $z \approx 6$, where
we know very little: only that some galaxies and QSOs already existed,
and that their energetic output may have just risen to the point of
reionizing the IGM \cite{Betal01,DCSM01}. From $z=6$ to
3, the age of the universe more than doubles, and its density decreases
more than five-fold.  We expect corresponding changes in the galaxy
population, but measuring this evolution will require large large,
systematic surveys for galaxies at $z > 4$ to compare with more than
1000 Lyman break galaxies (LBGs) now known at $z \approx 3$
\cite{SAGDP99}.  Color selection at $z \approx 5$ to 6 requires
deep imaging at $\lambda \geq 0.9\mu$m.  However, from the ground even
the brightest $z>5$ galaxies are barely detectable.  If there is no
evolution between $z=3$ and 6, we expect an $L^\ast_{UV}$ LBG to have
$m_{i} = 25.7$ at $z=5$, and $m_{z}=26.0$ at $z=6$. The HST Advanced
Camera for Surveys (ACS) will easily reach these limits thanks to the
dark sky and sharp image quality, which also provide greatly reduced
photometric error, incompleteness, and contamination due to confusion
with other (mostly foreground) objects. An $L^*$ LBG detected at $z=6$
by ACS will also be detectable in ultra-deep integrations with the
IRAC detector aboard SIRTF. Depending on the quality of the point-spread
function such detections will either be marginal (requiring prior
knowledge of source positions and possibly statistical co-addition 
of multiple galaxies for a secure detection) or clear cut
(if the PSF is as good as measured in pre-flight tests).
Figure 1 illustrates the expected yield of $z \sim 5.5$ LBGs from the
GOODS survey from a simple extrapolation of the $z=3$ luminosity function. 
We expect several hundred to of order 1000 candidates.

\begin{figure}[h!]
\begin{center}
\includegraphics[width=.7\textwidth]{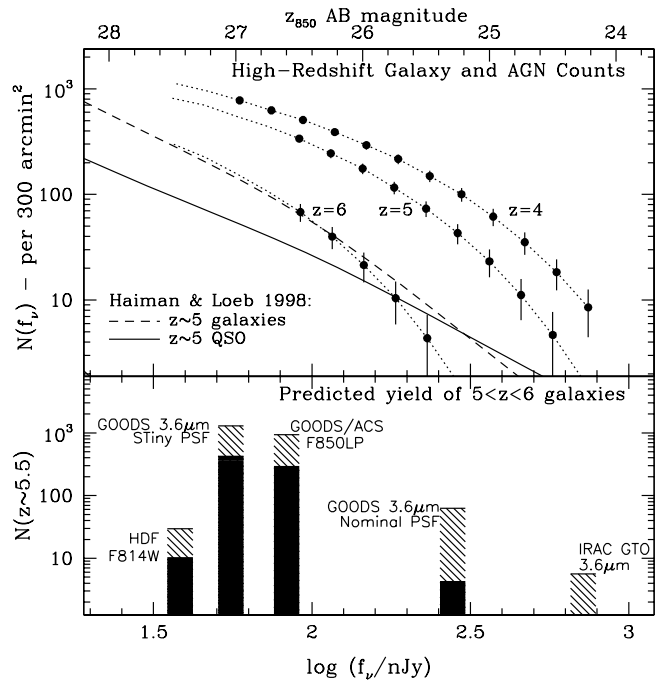}
\end{center}
\caption[]{
{\it Top:} Number of sources per unit redshift per logarithmic flux
interval anticipated in the GOODS survey region.  Solid and dashed
lines show the average number of 
galaxies and QSOs, respectively, per unit redshift $5 < z < 10$
from Haiman \& Loeb (1998).
Dotted lines show a model based on the measured luminosity
function of LBGs at $z \sim 3$, with $L^*$ scaled with redshift
as $(1+z)^{-3/2}$, i.e., following the Press-Schechter relationship
for halo mass.
Error bars represent Poisson statistics on galaxies recovered at
$z \sim 4, 5,$ and $6$ using simple $B$, $V$, and $i$--dropout
criteria.  Fluxes have been scaled to equivalent $z$-band fluxes
using a typical LBG SED.
{\it Bottom:}
Total galaxy counts at
$z = 5.5 \pm 0.5$ vs.\ the $10\sigma$ limiting depths (in nJy)
of various surveys, for a non-evolving LBG luminosity function (shaded bars)
and for one with the evolution adopted in the upper panel (solid bars).
Under these assumptions the HDF should have had 10-30 $z \sim 5.5$
galaxies, consistent with tentative identifications via
photometric redshifts.  The GOODS ACS survey will yield 300-1000
secure identifications.  Estimates of high-$z$ galaxy numbers are also
shown for the GOODS {\it SIRTF} 25-hour depth
survey, under two assumptions that bracket the range of expected
sensitivities due to uncertainties in the on--orbit IRAC PSF.
Predicted numbers (nominal PSF) are shown also for the planned
(0.3 deg$^2$) 3-hour IRAC GTO survey of the Groth strip. A cosmology
with $\Omega_M, \Omega_\Lambda, h = (0.3, 0.7, 0.65)$ is adopted.
}
\end{figure}

\section{Lyman-break Galaxies in the NGST Era}

The Next Generation Space Telescope, slated for launch before 2010, 
is expected to be a $\sim 6.5$ m passively-cooled telescope with 
exquisite sensitivity from 0.6 to 28 $\mu$m. Three focal-plane instruments
are planned: an optical/near-IR camera with a $4^\prime \times 4^\prime$
field of view, a low-resolution ($R \sim 100-1000$) near-IR spectrograph
(to be built by ESA), and a mid-IR camera/spectrograph. At 4.5 $\mu$m,
an exposure time of 5 seconds will be required to match the 50-hour
depth of IRAC GOODS images. Presuming that observations from SIRTF, ACS
and WFC3 are successful, by the time NGST launches galaxy populations
at $z < 6$ should be reasonably well understood. However, the fact
that reionization occurs at $z \gtrsim 6$ suggests that the earliest
luminous structures formed at still higher redshift. 

By redshift $z=3$ the massive stars that reionized the universe (if
indeed stars were responsible) exist only as undetectable remnants.
However, if early generations of stars formed with a Salpeter or Scalo
initial mass function (IMF), lower-mass cousins of the ionizing sources
would still be on the main sequence at $z=3$, and would perhaps reside
in Lyman-break galaxies. Existing optical and near-IR observations of
LBGs yield constraints on stellar populations that can be used to
explore this possibility. Papovich, Dickinson \& Ferguson \cite{PDF01} 
studied a sample of
spectroscopically-confirmed LBGs from the Hubble Deep Field North (HDF)
in the redshift range $2.0 \lesssim z \lesssim 3.5$.  The UV-optical
data were drawn from WFPC-2 observations, and the infrared from NICMOS
J and H-band observations and from $\rm K_s$-band observations with the
infrared imager IRIM at the KPNO 4m Mayall telescope
\cite{Dickinson98p219}.  Stellar-population models from 2000 version
of the Bruzual-Charlot \cite{BC93} code were fit to 31 galaxies, 
varying metallicity,
e-folding timescale $\tausf$, age, IMF (Salpeter, Miller-Scalo, Scalo),
extinction, and extinction law \citealp{CABKKS00,CCM89}. The
geometric mean of the best-fit ages for the sample is 0.12 Gyr for the
solar metallicity case.  Thus a typical galaxy observed at $z = 3.0$
would have ``formed'' at $z=3.15$.  Shapley et al. 
\cite{SSADGP01} analyzed
$G, {\cal R}, J,$ and $K_s$ groundbased photometry for a sample of %81
galaxies with spectroscopic redshifts $2.2 < z < 3.4$.
The published paper reports results for the best-fit continuous
star-formation models ($\tausf = \infty$) to the 74 galaxies for which
acceptable fits were obtained. The median best-fit age for this sample
is 0.32 Gyr, implying a formation redshift $z=3.4$ for a typical galaxy
observed at $z=3$.

\begin{figure}[h!]
\begin{center}
\includegraphics[width=.99\textwidth]{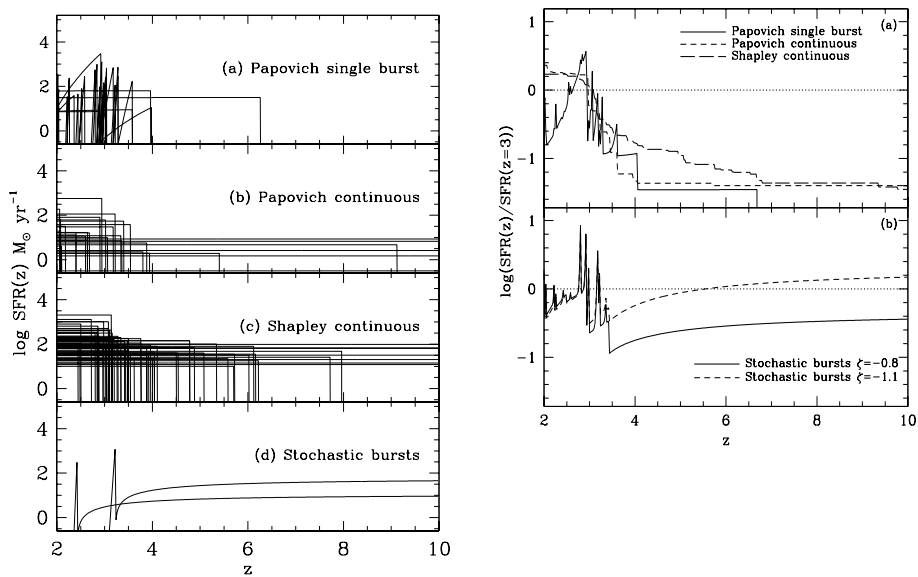}
\end{center}
\caption[]{{\it Left:}
Star-formation rate vs. time for individual galaxies, as inferred from
the SED models. The top panel shows the best-fit models with
exponentially declining SFR.
Panel (b) shows the star-formation histories from 
continuous star-formation models from \cite{PDF01} characterized by a
stellar mass $M$ and an age.
Panel (c) shows the same kind of continuous star-formation
model for the Shapley et al. \cite{SSADGP01} sample. Panel (d) shows two examples
of the stochastic burst model described in the text applied to galaxies
97 and 1115 in the PDF01 sample.
{\it Right:}
Global star-formation rate vs. time for all of the models, normalized
to the rate at $z=3$. In the top panel, the solid curve is for the
PDF01 $\tau$ models.  The short-dashed curve is for their continous
star-formation models. The long-dashed curve is for the Shapley et al. 
\cite{SSADGP01} continuous star-formation models.
The bottom panel shows the star-formation rate vs. time for the
stochastic burst models with a Salpeter IMF (solid) and a top-heavy
IMF with $x=0.5$ (dashed).
Star-formation rates are normalized to the mean in the range $2.5 < z < 3.5$.
}
\end{figure}

In Figure 2a, we show the star-formation histories derived for each
galaxy in the two samples, under various assumptions. The top panel
shows exponentially-declining models. The second and third panels show
models with a constant star-formation rate (where only the age, extinction,
and total stellar mass are free parameters). 
In the exponentially decaying models
only one out of the 31 galaxies would have been present at $z = 6$. In
the oldest continuous-star-formation models from \cite{PDF01}, 
six out of 31 or 19\%
would have been present at $z=6$.  
The Shapley et al. \cite{SSADGP01} models imply that only 17\% of the
galaxies were present at $z=6$.

Models with two distinct episodes of star formation allow more star
formation at higher redshift. Papovich et al. \cite{PDF01} fit maximally-old models to
their LBG sample, deriving constraints on the mass of an old population
that formed with a Salpeter IMF in an instantaneous burst at $z =
\infty.$ This model quantifies how much stellar mass can be hidden
``underneath the glare'' of the young population that dominates the
UV/Optical radiation from each galaxy.  However, the star-formation
rate predicted at $z=6$ from such maximally old components is zero,
because all star-formation happened at higher redshift. It is more
likely that starbursts induced by mergers are spread out over some
range of redshift and do not occur in all galaxies simultaneously.  If
the older burst in the LBGs is put at redshift lower than $z = \infty$,
the mass in the burst must be lower. Rather than fit a whole suite of
models of different burst redshifts, we can, to a good approximation,
scale the allowable mass in the old component by a power-law fading
model.  By fitting the B-band luminosity vs.  time for $10^7 < t < 2
\times 10^9 \,\rm yr$, we find $L_{\rm B} \propto t^{-0.8}$ for a
Salpeter IMF for an instantaneous burst in the Bruzual \& Charlot
solar-metallicity models.  If each galaxy had an instantaneous
probability $P(z)$ of forming stars at redshift $z$, and a typical
burst had a duration $\Delta t$ the average SFR from an ensemble of
such galaxies would be $\xi(z) = M(z)P(z)/\Delta t$,
where $M(z)$ is average mass formed in each burst and $\Delta t$ is the
average duration of each burst.  For simplicity we
adopt a constant $P(z)$ from $z=10$ to the observed LBG redshift
$z_{\rm obs}$.

Figure 2d shows the SFR vs. redshift implied by such a stochastic model
for two individual galaxies (numbers 97 and 1115) in the PDF01 sample.
The low-redshift spikes in the star-formation rate correspond to the
young component that dominates the light at the observed redshift; the
star-formation progressing to higher redshift represents the mean for
an ensemble of stochastic bursts. Obviously any single galaxy would
simply show two spikes of star formation for this kind of model, but if
we consider such a galaxy as a proxy for millions of others, the
star-formation history shown in the figure represents the maximal rate
of star-formation due to stochastic bursts as a function of redshift.

The constraints become clearer if we consider the entire
sample of galaxies.  Figure 3 shows the evolution of $\SFRDz$ with time
relative to that $z = 3$ computed by summing up the models shown in the
previous figures. The top panel shows the smooth star-formation
histories.  For these cases the inferred co-moving density of star
formation declines dramatically from $z=3$ to higher redshift.  Even if
we put the maximum mass allowed in stochastic-starbursts at redshifts
$z > z_{\rm observed}$, the star-formation rate at $z=6$ is still a
factor of 3 below that at $z=3$, as shown by the solid curve in Fig.
3b.

Adopting $\fesc = 0.1$, the required density of star-formation for
reionization in the Madau, Haiman \& Rees \cite{MHR99} model is a factor of 1.3 times higher
than the dust-corrected $\SFRD$ at $z \sim 3$ measured by
\cite{SAGDP99}.  In contrast, the star-formation rates inferred from
the SED fits imply a sharp decrease in $\SFRD$ between $z=3$ and
$z=6$.  The problem becomes even more severe if a significant fraction
of the baryons are already collapsed into minihalos at the time of
reionization.  In this case the required number of ionizing photons
increases by a factor of 10-20 \cite{HAM01}, and all models fall short
even if $\fesc = 1$.

There is another, perhaps more serious,
problem with the star-formation histories derived so far:
all the models imply a dramatic decline in star-formation by $z=4$.
But the observed LBG rest-frame UV luminosity functions are very similar
at $z = 3$ and $z=4$, and the integrated star-formation
rates derived therefrom differ only by a factor of $1.1 \pm 0.4$
\cite{SAGDP99}. Thus the star-formation {\it histories} derived from the $z=3$
LBGs are in direct conflict with the star-formation {\it rates} derived
for the $z=4$ LBGs. 

More star formation can be hidden in bursts if the bursts fade faster.
If the fading exponent over an age range $10^7$ -- $2 \times 10^9$ yr
is $\zeta$, the additional increase in stellar mass that can be hidden
relative to a Salpeter IMF is roughly $t_7^{-\zeta-0.8}$, where $t_7$
is the age in units of $10^7$ yr. At an age of 1 Gyr a factor of four
more stellar mass can be hidden if $\zeta = -1.1$ than if $\zeta =
-0.8$. Indeed, changing the fading exponent to $\zeta = -1.1$ is
sufficient to make the allowed star-formation rate at $z=6$ equal to
the star-formation rate at $z=3$. The global star-formation rate from
such a model is shown as a dashed curve in Fig. 2b.  If the IMF is a
powerlaw $\phi(M)dM \propto M^{-(1+x)}$, a fading exponent $\zeta =
-1.1$ requires an IMF slope $x = 0.5$ compared to the Salpeter value $x
= 1.35$ (for an instantaneous-burst solar-metallicity stellar
population). A steeper fading slope $\zeta = -1.2$ (corresponding to an
IMF slope $x = 0.3$) is needed to bring $\SFRD$ at $z=4$ to within a
factor of 1.3 of that at $z=3$. Lower metallicities require even more
top-heavy IMFs.  Options other than varying the IMF are of course
possible (e.g.  evolved stellar populations could be hidden by dust
that builds up over timescales of $10^8$ to $10^9$ yrs). However, the
requirement for faster-than-Salpeter fading is robust. Furthermore, the
fading must be even faster if galaxies on average have more than two
burst episodes.

Constraints on the star-formation histories of LBGs will improve greatly
over the next few years with the advent of SIRTF and ACS. Observations
with these instruments will further narrow the parameter space available
for bursty or episodic star-formation. Such observations will set the stage for
the detailed exploration of galaxy formation in the
``pre-reionization'' era at $z > 6$ with NGST.

We would like to thank our collaborators on the HDF observations for
their many contributions to this work. We thank Jennifer Lotz and
Mauro Giavalisco for valuable discussions. Support for this work was
provided by NASA through grant GO07817.01-96A from the Space Telescope
Science Institute, which is operated by the Association of
Universities for Research in Astronomy, under NASA contract
NAS5-26555.

\bibliographystyle{apj}
\renewcommand\bibsection{\section*{\refname}}
\addcontentsline{toc}{section}{References}
\bibliography{apjmnemonic,bib}

%INDEX%%%%%%%%%%%%%%%%%%%%%%%%%%%%%%%%%%%%%%%%%%%%%%%%%%%%%%%%%%%%%%%
% Please check with the editor of your book whether he plans to
% include a "mutual" subject index - if so, please code your entries
% in the standard syntax. For your own purposes you may print your
% "personal" index by using the following commands:
%
%\clearpage
%\addcontentsline{toc}{section}{Index}
%\flushbottom
%\printindex
%%%%%%%%%%%%%%%%%%%%%%%%%%%%%%%%%%%%%%%%%%%%%%%%%%%%%%%%%%%%%%%%%%%%%

\end{document}